\documentclass[3p,twocolumn,sort&compress]{elsarticle}
\usepackage{graphicx}
\usepackage{amssymb}
\usepackage{amsthm}
\usepackage{epstopdf}
\usepackage{dcolumn}
\usepackage{bm}

\begin{document}

\begin{frontmatter}

\title{Carbon release by selective alloying of transition metal carbides}
\author[phy,cadd1]{Mikael R{\aa}sander\corref{cor1}}
\ead{mrasander@ucdavis.edu}

\author[che,cadd2]{Erik Lewin}

\author[che]{Ola Wilhelmsson}

\author[phy]{Biplab Sanyal}

\author[phy]{Mattias Klintenberg}

\author[phy]{Olle Eriksson}

\author[che]{Ulf Jansson}

\cortext[cor1]{Corresponding author. Tel.: +1-530-752-2297}
\address[phy]{Division for Materials Theory, Department of Physics and Astronomy, Uppsala University, P.O. Box 516, 75120 Uppsala, Sweden}
\address[che]{Department of Materials Chemistry, Uppsala University, P.O. Box 538, 75121 Uppsala, Sweden}
\fntext[cadd1]{Presently at Department of Chemistry, University of California Davis, One Shields Avenue, 95616 Davis, CA, USA.}
\fntext[cadd2]{Presently at Nanoscale Materials Science, Empa - Swiss Federal Laboratories for Materials Science and Technology, D{\"u}bendorf, Switzerland.}

\date{\today}

\begin{abstract}
We have performed first principles density functional theory calculations on TiC alloyed on the Ti sublattice with 3d transition metals ranging from Sc to Zn. The theory is accompanied with experimental investigations, both as regards materials synthesis as well as characterization.
Our results show that by dissolving a metal with a weak ability to form carbides, the stability of the alloy is lowered and a driving force for the release of carbon from the carbide is created. During thin film growth of a metal carbide this effect will favor the formation of a nanocomposite with carbide grains in a carbon matrix. The choice of alloying elements as well as their concentrations will affect the relative amount of carbon in the carbide and in the carbon matrix. This can be used to design the structure of nanocomposites and their physical and chemical properties. One example of applications is as low-friction coatings. Of the materials studied, we suggest the late 3d transition metals as the most promising elements for this phenomenon, at least when alloying with TiC.
\end{abstract}


\end{frontmatter}
\section{\label{sec:introduction}Introduction}
\begin{figure}[t]
\begin{center}
\includegraphics[width=7.0cm]{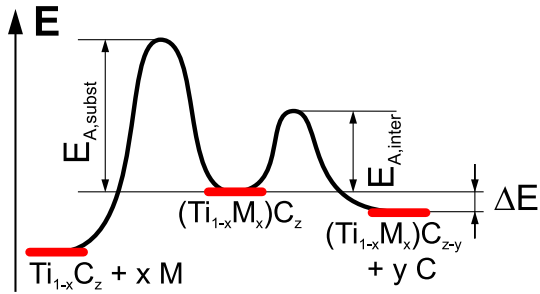}
\caption{(Color online) Illustration of possible decomposition routes of the metastable ternary solution $\rm {Ti}_{1-\rm {x}}\rm {M}_{\rm {x}}\rm {C}_{\rm {z}}$. Note the definition of the energy for carbon release from the carbide phase, $\Delta E$.}
\label{fig:barrier}
\end{center}
\end{figure}

Transition metal carbides (TMC) have for a long time been the focus of extensive research, due to their interesting physical properties such as high melting points, hardness and conductivity.  Many TMC are also used as components in hard metal tools or as wear-resistant coatings on such tools.\cite{toth} A well known technique to deposit thin coatings of TMC is magnetron sputtering and one of the most studied TMC coating materials is TiC. With higher carbon contents the sputtered coatings usually form nanocomposites with nanocrystalline carbide grains in a carbon matrix or tissue phase, i.e. nc-TiC/a-C.\cite{refA} The properties of the TMC nanocomposite strongly depends on 
 several factors such as the carbide grain size, thickness of the matrix phase and the relative distribution of carbide and matrix. Usually the growth process in sputtering is carried out far from equilibrium and it is therefore possible to alloy a carbide film with additional transition metal atoms by a substitutional solution on the Ti sublattice. The solubility of such metals in the sputtered coatings can be far higher than allowed at thermodynamical equilibrium, see e.g. Refs. \cite{refBa,refBb,refBc,refB4}, hence producing metastable phases. During annealing or under external pressures, there will be a driving force for the metastable TMC coating to decompose into more stable phases. A schematic view of this concept is shown in Fig. \ref{fig:barrier} illustrating the decomposition process  of a hypothetical metastable Ti$_{1-{\rm x}}$M$_{\rm x}$C$_{\rm z}$ phase where the weak carbide forming metal M is substitutionally dissolved into the TiC structure. The thermodynamically most favourable decomposition route is always to remove the metal from the carbide structure by solid state diffusion and nucleation to form a metal-rich phase (left reaction route in Fig. \ref{fig:barrier}). However, this requires substitutional diffusion of the metallic species which always has a much higher activation energy than interstitial diffusion. An alternative and kinetically more favourable route is therefore interstitial diffusion of carbon towards the surface of the grains and  a subsequent formation of free carbon (right reaction route in Fig. \ref{fig:barrier}).  Of course, this requires that it is energetically favorable for the ternary carbide to release some of its carbon. The exact value of the activation energies, $E_{A}$,  in the Ti$_{1-{\rm x}}$M$_{\rm{x}}$C phase will depend on the concentration of the metal M, however, literature data are known for the corresponding energies in pure TiC.\cite{refC} For this case, activation energy for interstitial diffusion of carbon atoms has been determined to 4.15 eV which should be compared with substitutional diffusion of Ti with an activation energy of 7.65 eV.\cite{refC} We note that even though the activation energies for diffusion of the metal M and carbon in the ternary solid solutions are not known, there is no reason to assume that the activation energy for substitutional diffusion of the metal M will be lower than the activation energy for carbon diffusion in these systems.
\par
The existence of two possible reaction routes will influence the growth of ternary, metastable  carbides. Firstly, it suggests that a solid solution of a weak carbide forming metal into e.g. TiC should favour the formation of free carbon between the grains and on the surface of the film. This has also been observed experimentally by e.g. Wilhelmsson et al. in Refs. \cite{refB4} and \cite{wilhelmsson} for the case of alloying of Al into TiC. This effect can also be used to dramatically form low-friction graphitic surfaces and design new types of low-friction materials.\cite{wilhelmsson} Secondly, it also suggests that a process window should exist where at high temperatures a metal-rich phase is formed (left reaction route in Fig. \ref{fig:barrier}), while at lower temperatures mainly carbon diffusion according to the right reaction route in Fig.~\ref{fig:barrier} will be found.
\par
In Ref. \cite{wilhelmsson}, we were able to demonstrate that metastable solid solutions of Ti$_{1-\rm {x}}$Al$_{\rm {x}}$C$_{\rm {z}}$ coatings could be deposited by magnetron sputtering and that they in agreement with theoretical predictions  indeed formed surface graphitic layers upon both an annealing step and in tribological contacts. The surface layer of graphite reduced the friction coefficient to 0.05-0.1 using steel as a counter surface. An interesting observation is that coatings with a dissolved weak carbide forming metal is less hard but can exhibit similar wear properties compared to a pure TiC coating. \cite{Wear1,Wear2,refB4}
\par
The concept of dissolving a weak carbide-forming metal into a TMC to modify the structure and properties of a coating can be expanded to a large number of carbides and wide range of dissolving metals. However, a systematic theoretical and experimental investigation of this effect has yet not been carried out. The stability of metal carbides has been studied by many authors, e.g. in Refs. \cite{refEa,refEb,refEc,refEd}. The most important result within the present context is that the stability of the transition metal carbide is reduced with increasing number of d electrons. In fact, metals such as Fe, Ni and Cu form no thermodynamically stable carbides and it is therefore likely that different types of low-friction coatings can be designed by a solid solution of such weak carbide forming elements.
\par
The aim with this study is to carry out a systematic investigation of the stability of a ternary solid solution of the metal M into TiC according to:
\begin{equation}\label{eq:carbform}
(1-\rm {x})\cdot\rm {Ti} + (1-\rm {y})\cdot\rm {C} + \rm {x}\cdot\rm {M} \longrightarrow \rm {Ti}_{1-\rm {x}}\rm {M}_{\rm {x}}\rm {C}_{1-\rm {y}}.
\end{equation}
The stability of this solid solution has been investigated by calculations of the energy of formation with respect to Ti, M and C in their corresponding reference states. In this paper M has been restricted to the 3d transition metals series from Sc to Zn.
Furthermore, the driving force for the out-diffusion of carbon under the formation of graphite has been calculated based on the energy difference between the phases in the reaction: 
\begin{equation}\label{eq:carbonform}
\rm {Ti}_{1-\rm {x}}\rm {M}_{\rm {x}}\rm {C} \longrightarrow \rm {Ti}_{1-\rm {x}}\rm {M}_{\rm {x}}\rm {C}_{1-\rm {y}} + \rm {y}\cdot\rm {C},
\end{equation}
where C is in the form of graphite. This is equivalent to the evaluation of $\Delta E$ in Fig. \ref{fig:barrier}. The result from the theoretical calculations have been compared with selected experimental data using  Ni and Cu as alloying metals. As will be shown the experimental results are in good agreement with the theoretical predictions. 
\par
The paper is arranged as follows: In Section \ref{sec:method} we will discuss the necessary theoretical and experimental background, in Section \ref{sec:theoretical results} we will discuss the theoretical results, in Section \ref{sec:experimental results} the experimental results are discussed and in Section \ref{sec:conclusion} we will summarize and draw conclusions.
\begin{figure*}[t]
\begin{center}
\includegraphics[width=7.0cm]{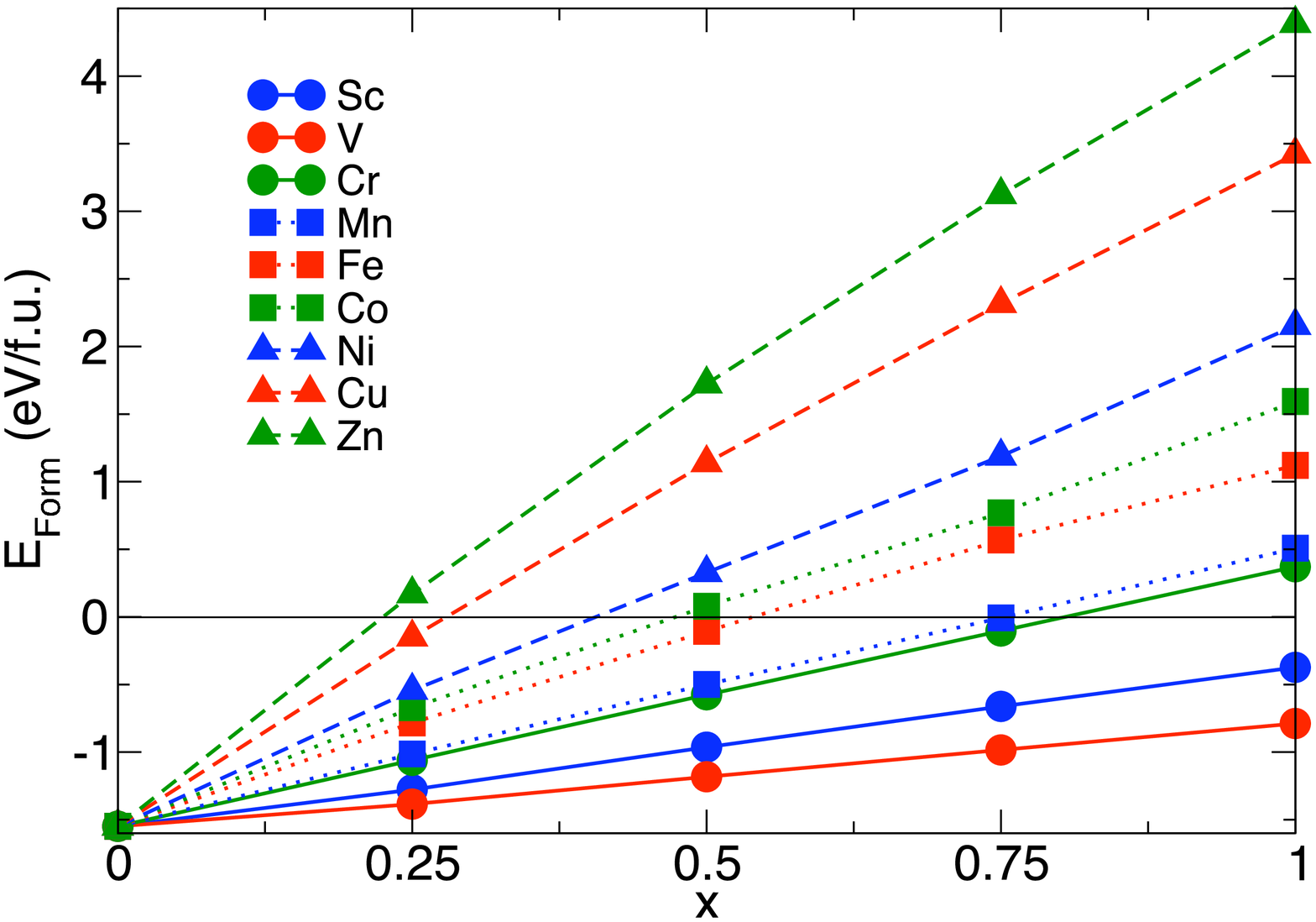}\qquad\includegraphics[width=7.0cm]{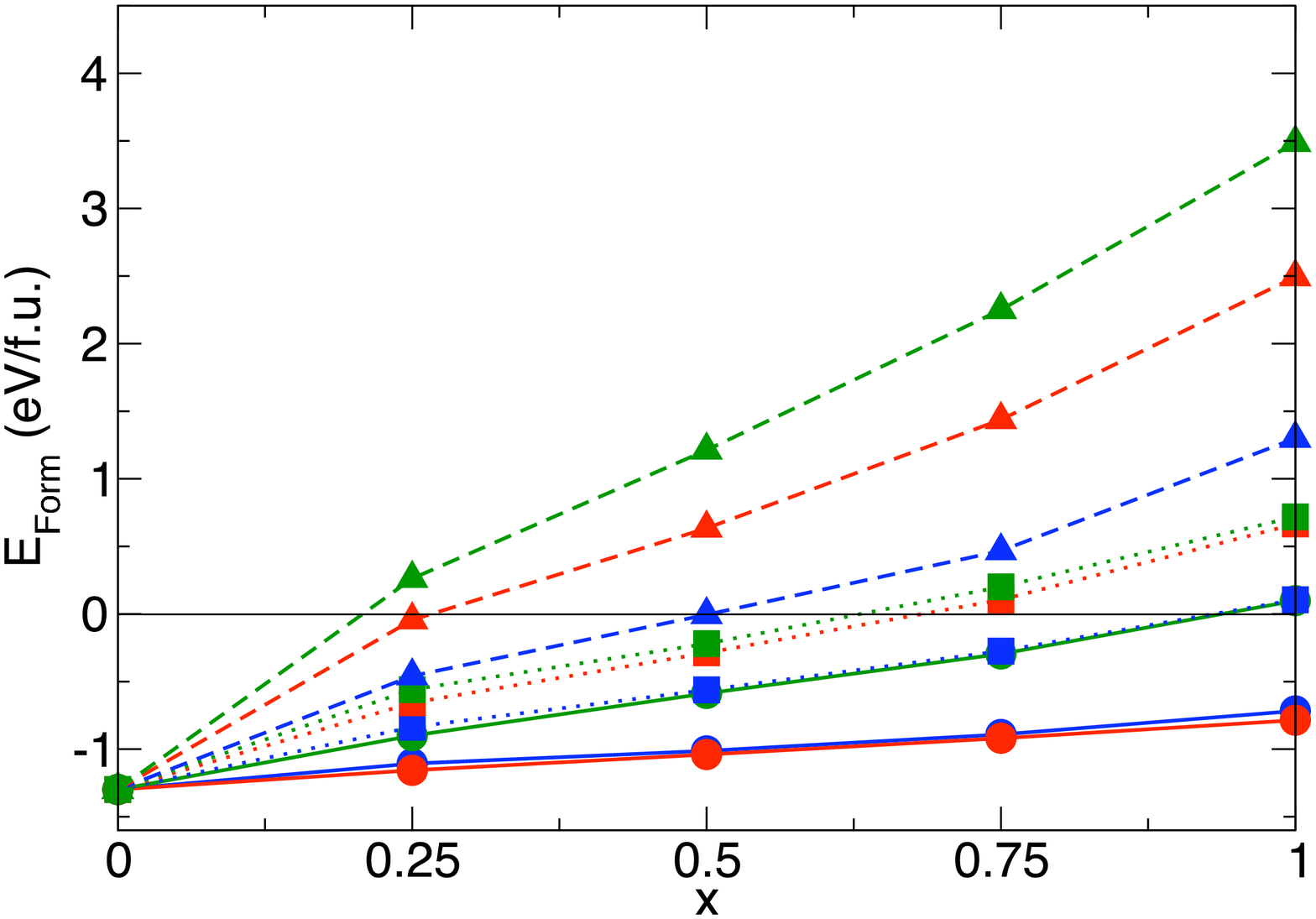}
\caption{(Color online) The formation energy, $E_{Form}$,  for the ternary solution $\rm {Ti}_{1-\rm {x}}\rm {M}_{\rm {x}}\rm {C}_{1-\rm {y}}$ according to Eq. (\ref{eq:formationenthalpy}) as a function of concentration, x, of the alloying element M for both full stoichiometric (y=0, left panel) and carbon deficient (y=0.25, right panel) phases. Energies are in units of eV/f.u., where f.u. stands for formula unit.\label{fig:formationenthalpy}}
\end{center}
\end{figure*}
\section{\label{sec:method}Method}
\subsection{Theory}
The TMC that we focus on in this study all have the B1 (NaCl) structure, i.e. a face centered cubic metal lattice with C atoms occupying the octahedral interstitial positions. This is the ground state structure of TiC for a large compositional range when it comes to the C concentration.\cite{toth,Tana,Tanb} For VC the B1 structure is also the ground state crystal structure but in a much more narrow compositional range.  For 3d transition metals from Cr to Zn there are essentially no chemically stable carbides, at least not in the B1 crystal structure. \cite{toth} However, since we are interested in the stability of alloyed TiC the B1 structure serves as a good reference crystal structure.
\par
In order to model the ternary TMC, the M atoms are substitutionally dissolved into TiC by substituting some of the Ti atoms on the metal lattice. There are a number of issues related to the treatment of such a system. 
Assuming that the substitution on the metal lattice will leave the C lattice unaffected, it is necessary to consider the effects of disorder on the metal lattice. Furthermore, since the TMC have relatively open structures, and the fact that the metals used in this study have different ability to form carbides,\cite{toth,refEa,refEb,refEc,refEd} it is likely that the structure and stability of these carbides will depend on the local environment in different parts of the system.  Recently there has been a study on the importance of incorporating local environment effects in these types of systems.\cite{Alling}  Therefore, it is preferable to use a method that takes both disorder and local environment effects into account. Here we treat the problem of disorder within the supercell approach, where we have considered special quasi random structures (SQS) that mimics the disorder of a random lattice.\cite{Zunger} The supercells that have been used in the calculations all contain 24 atomic positions on both the metal and the carbon sublattices. This means that in order to have 25~\% of the Ti atoms substituted for another metal atom, the supercell will contain 18 Ti and 6 M atoms. For 50~\% substitution, there will be 12 atoms of each species. For the carbon sublattice, the relation between C atoms and C vacancies works in a similar fashion as for the metallic sublattice. Similar structures have been used in, for example, Ref. \cite{Alling}.
\par
The calculations have been performed within density functional theory \cite{HohenbergandKohna,HohenbergandKohnb} using the projector augmented wave method \cite{bloechl} as it is implemented in the Vienna {\it ab initio} simulation package. \cite{KresseandFurth,KresseandFurthb} We have used the generalized gradient approximation for the description of the exchange and correlation functional.\cite{perdewandwang91}
The cut-off energy for the plane wave basis set was set to 600 eV in all calculations, and we have used the special k-points method of Monkhorst and Pack\cite{MonkhorstandPack} and made sure that the calculations have been converged with respect to the number of k-points. Spin-polarized calculations have been performed for systems containing the metals Mn, Fe, Co, and Ni, however, results related to the magnetic properties of these phases will be presented elsewhere.

\subsection{Experimental}
To verify the theoretical results, two sets of Ti-M-C films were deposited by non-reactive dc magnetron sputtering in an ultra-high vacuum chamber (base pressure below 1.0$\cdot10^{-7}$~Pa) using an Ar-discharge of 4.0$\cdot10^{-1}$~Pa. Elemental 2" targets of Ti (99.995\% purity), M ($>$99.99\% purity) and C (99.999\% purity) were used. In the present study Ni and Cu were used as alloying elements. The films were deposited on substrates of Si(100). The films were characterized with X-ray photoelectron spectroscopy (XPS) and X-ray diffraction (XRD). XPS was performed using a Physical Electronics Quantum 2000 spectrometer and samples were sputtered with 200 eV Ar$^+$-ions prior to analysis to remove surface oxides. For all XPS analysis monochromatic Al K $\alpha$ radiation was used. Total composition of the samples was also determined using XPS data. XRD was performed using a Philips XÍPert MRD diffractometer, and grazing incidence (GI) scans were performed with an incident angle of 2 degrees. Annealing was performed in a vacuum furnace at pressures below $7.0\cdot10^{-5}$~Pa.
\section{\label{sec:theoretical results}Theoretical Results}
We begin the presentation of our results by an analysis of the energy of formation, $E_{Form}$, for the reaction given in Eq. (\ref{eq:carbform}), evaluated according to
\begin{eqnarray}\label{eq:formationenthalpy}
E_{Form} &=& E(\rm {Ti}_{1-\rm {x}}\rm {M}_{\rm {x}}\rm {C}_{1-\rm {y}}) \nonumber  \\
 & & -(1-\rm {x})\cdot E(\rm {Ti})  -\rm {x}\cdot E(\rm {M}) \\
 & & -(1-\rm {y})\cdot E(\rm {C}),\nonumber 
\end{eqnarray}
where $E(\rm {Ti}_{1-\rm {x}}\rm {M}_{\rm {x}}\rm {C}_{1-\rm {y}})$ is the total energy of the ternary carbide $\rm {Ti}_{1-\rm {x}}\rm {M}_{\rm {x}}\rm {C}_{1-\rm {y}}$ and $E(\rm {X})$ is the total energy of X, with X being either Ti, C, or M. The energies of Ti, C and of the metal M is with respect to each elements reference state in the bulk. The results are presented in Fig.~\ref{fig:formationenthalpy} for both stoichiometric (y=0) and sub-stoichiometric (y=0.25) phases.
The calculations clearly show that the most stable carbide phase is pure TiC, and that by increasing the M concentration, x, the formation energy increases. A positive formation energy corresponds to an unstable situation since the constituents will prefer to be separated in their corresponding single element phases. Note that there are only three of the binary carbide systems (TiC, ScC, and VC) with a negative formation energy, signifying a good carbide forming ability.
This is also in agreement with experimental data where only Ti, Sc, and V form stable monocarbides with the B1 structure. As can be seen by comparing left and right parts of Fig. \ref{fig:formationenthalpy}, the sub-stoichiometric ScC$_{0.75}$ is more stable than the stoichiometric ScC, which is also in agreement with experimental literature.\cite{refD,refD2} Our results also show that by alloying TiC with Sc and V on the metal sublattice results in negative formation energies for all values of the M content x. Overall, the ternary carbides show a monotonous behavior of the formation energy with respect to the alloying component. 
An interesting parameter in Fig. \ref{fig:formationenthalpy} is the slope of the curve. The steeper the slope, the more unfavorable is a solid solution, i.e. the less stable is the ternary solution. As can bee seen in Fig. \ref{fig:formationenthalpy}, the variation in the slope is directly a function of the number of valence electrons in the system, which is in agreement with previous theoretical studies.\cite{refEa,refEb} Zn and Cu which form no carbides give clearly the most unfavorable solid solutions.\par
By comparing the stoichiometric and non-stoichiometric carbides we conclude that having a lower carbon concentration in the structure reduces the formation energy in general making the sub-stoichiometric ternary phases less unstable than the corresponding stoichiometric system as is readily shown in Fig. \ref{fig:formationenthalpy}. This effect can be explained by the fact that in transition metal carbides the chemical bonding is known to be primarily due to strong metal-carbon bonds and a somewhat weaker metal-metal binding.\cite{Gelatt} 
An in-mixing of late transition metals disrupts the strong covalent bonding between e.g. Ti and C atoms in TiC, since the M-C bond is less favorable compared to the Ti-C bond, which increases the total energy of the system.  If we now consider an increasing amount of vacancies on the C lattice the importance of the metal-metal bond becomes more and more pronounced, simply because the number of C atoms to bind to is reduced. If the extreme case of 100\% C vacancies is considered we are left with only metal-metal bonds in a face centered cubic lattice. So by alloying TiC with other transition metals, the strong metal-carbon bond becomes less favorable creating a tendency for the creation of C vacancies. The sub-stoichiometric phases on the other hand, while having a lower number of unfavorable metal-carbon bonds in the system, are also stabilized due to an increased metal-metal bonding in these phases. This change in behavior of the bonding has also been shown to be reflected in the magnetic exchange interactions between Fe atoms in ternary Ti-Fe-C.\cite{Ti-Fe-C} 
\par
\begin{figure}[t]
\begin{center}
\includegraphics[width=7.0cm]{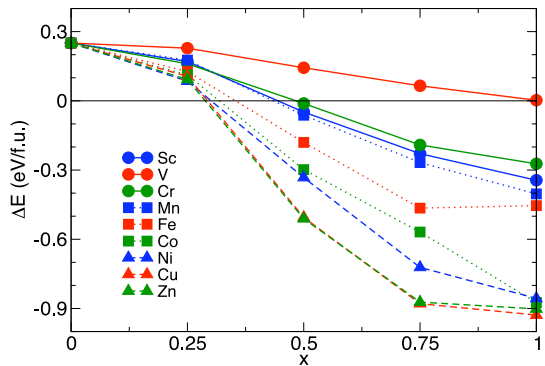}
\caption{(Color online) The energy difference, $\Delta E$, for the segregation of carbon from the ternary carbides $\rm {Ti}_{1-\rm {x}}\rm {M}_{\rm {x}}\rm {C}$ as defined in Eq. (\ref{eq:energydiff}) as function of M concentration, x.  Energies are in units of eV/f.u., where f.u. stands for formula unit.}
\label{fig:ediff}
\end{center}
\end{figure}
\begin{figure}[t]
\begin{center}
\includegraphics[width=7.0cm]{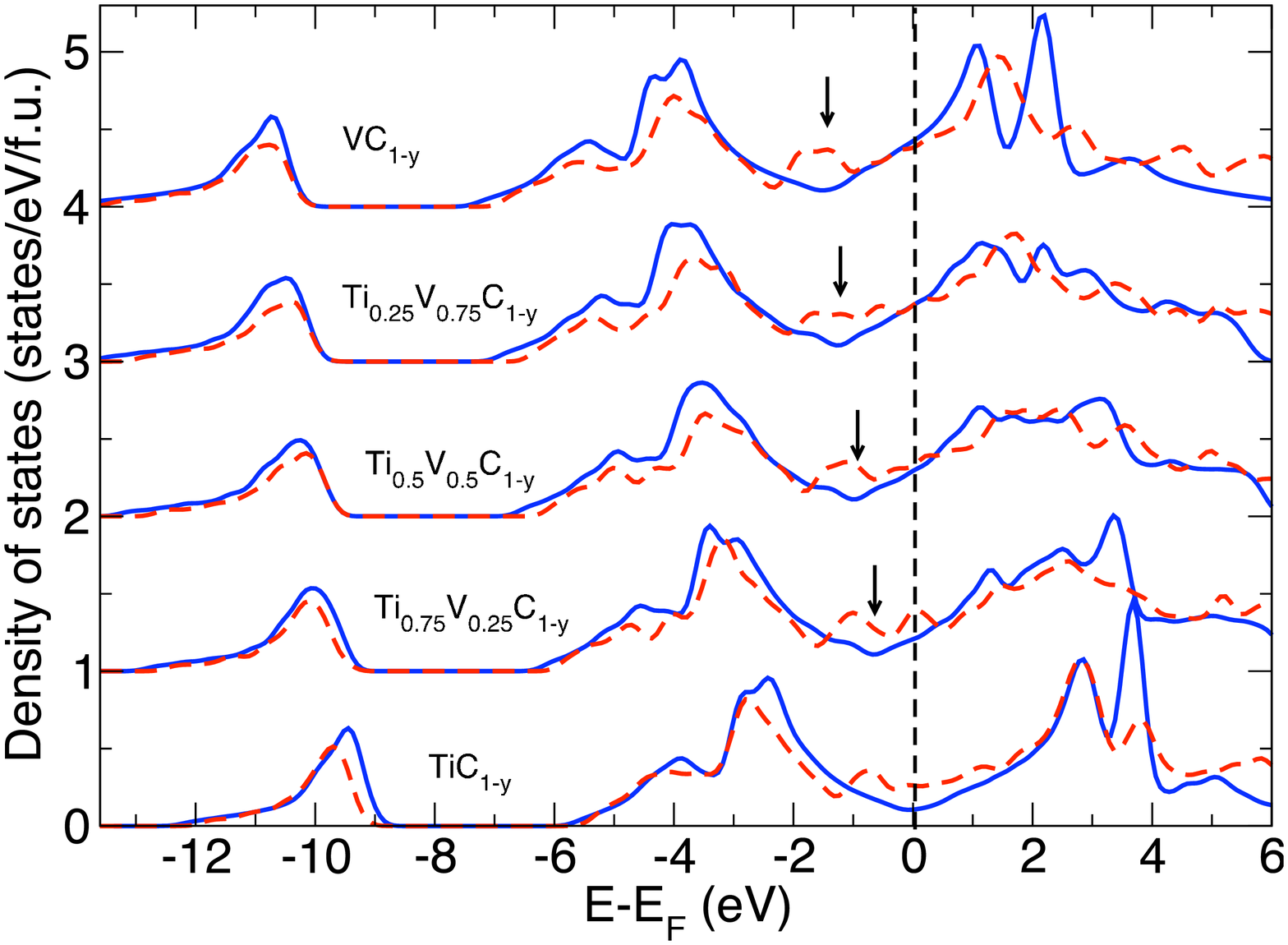}
\caption{(Color online) The calculated density of states of the Ti$_{1-\rm {x}}$V$_{\rm {x}}$C$_{1-\rm {y}}$-phases, for both y=0 (blue full line) and y=0.25 (red dashed line). The dashed vertical line represents the position of the Fermi level, $E_{F}$. The arrows represent the positions of the division between bonding and anti-bonding states within each phase.\cite{Grechnev}  For each step in x the DOS has been shifted with 1 states/eV/f.u., where f.u. stands for formula unit.}
\label{fig:DOStivc}
\end{center}
\end{figure}
\begin{figure*}[t]
\begin{center}
\includegraphics[height=6.0cm]{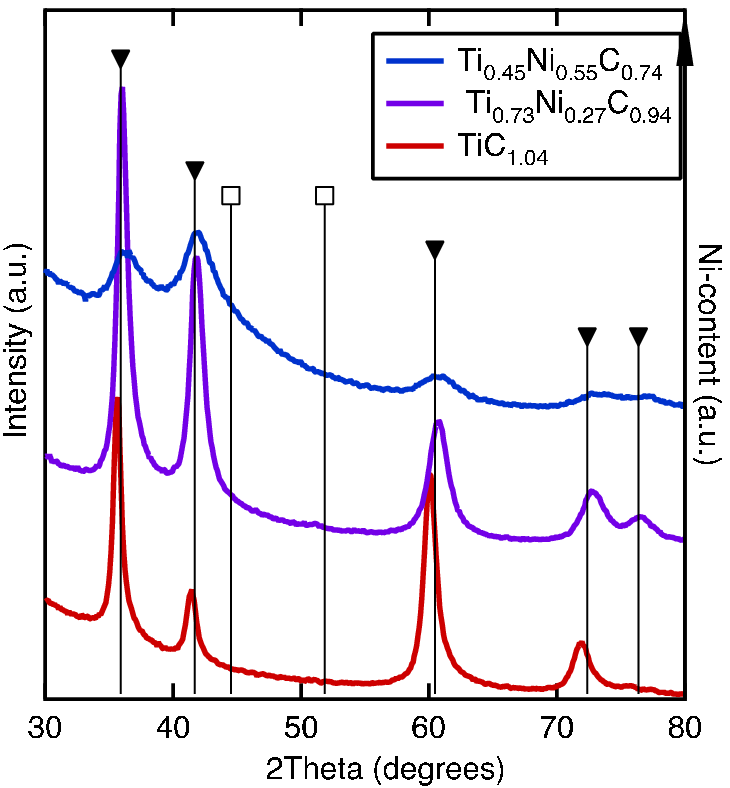}\qquad\includegraphics[height=6.0cm]{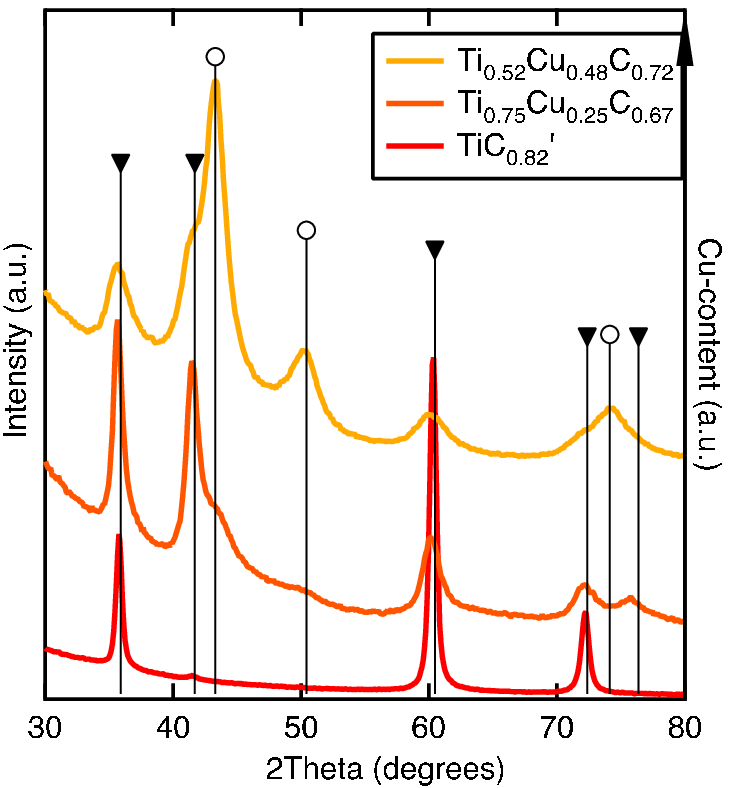}\\
\vspace{0.5cm}
\includegraphics[height=6.0cm]{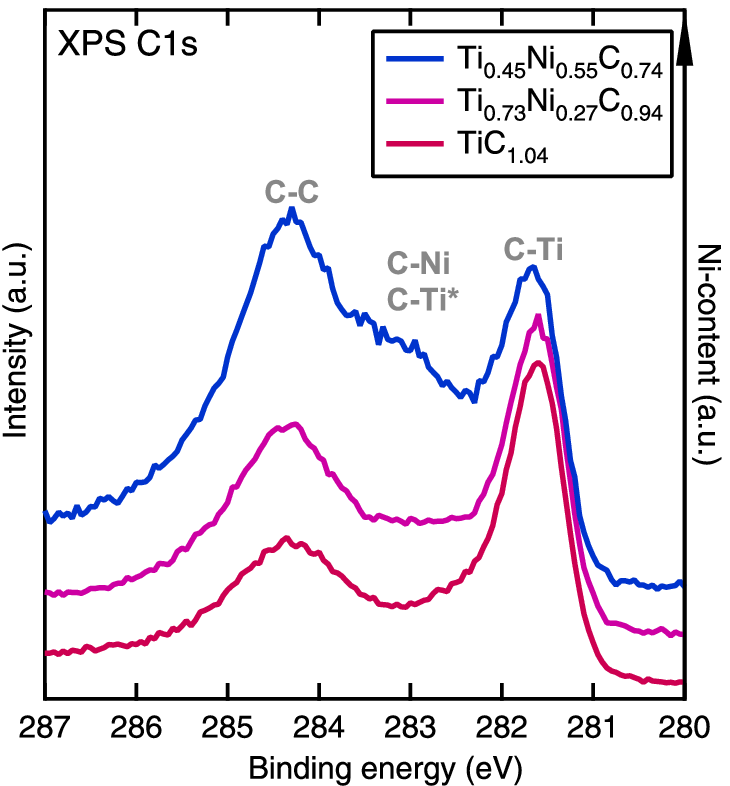}\qquad\includegraphics[height=6.0cm]{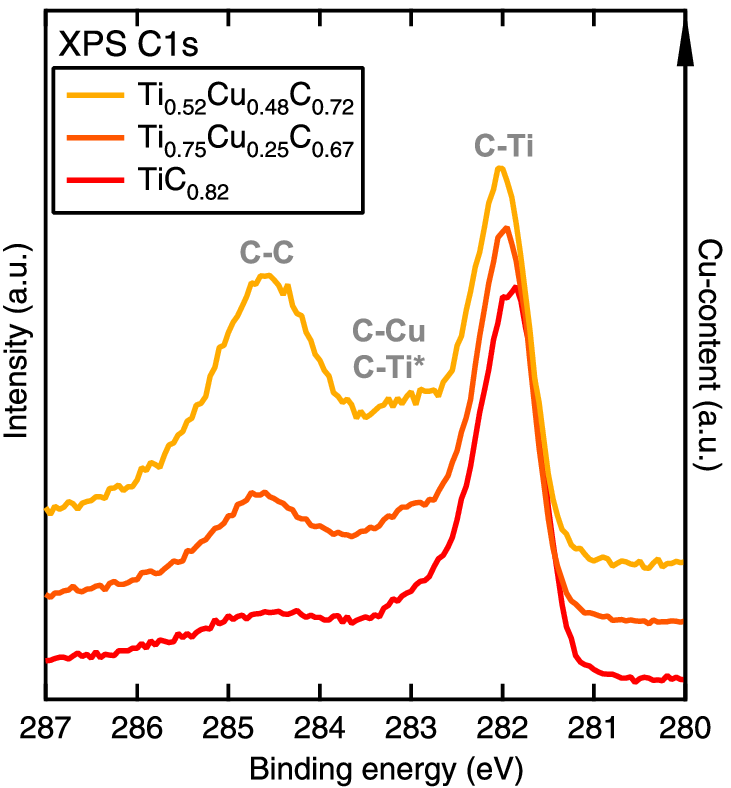}
\caption{(Color online) Diffractograms of Ti-Ni-C (upper left figure) and Ti-Cu-C (upper right figure) thin films deposited under identical conditions. Reference positions for TiC (triangles), Ni (squares) and Cu (circles) shown with markers.\cite{Morris,Swanson} C1s XPS spectra from two series of Ti-M-C nanocomposite thin films (lower panels). Ti-Ni-C samples with comparable total carbon content (lower left figure) and Ti-Cu-C samples, also with comparable carbon content (lower right figure). Spectra have been normalized to the C-Ti peak. As the Ni or Cu content increases the relative amount of C-C bonds also increases.}
\label{fig:tinicXPS}
\end{center}
\end{figure*}
We now turn our attention to the possibility of C to leave the system in the form of graphite as a response to the alloying with the metal M.  The relevant property to consider from the theory is in this case the energy difference, $\Delta E$, which is illustrated in Fig. \ref{fig:barrier}. This energy is given by the difference between the phases in the left and right parts of Eq.~(\ref{eq:carbonform}):
\begin{eqnarray}\label{eq:energydiff}
\Delta E &=& E(\rm {Ti}_{1-\rm {x}}\rm {M}_{\rm {x}}\rm {C}_{1-\rm {y}}) + \rm {y}\cdot E(\rm {C}) \nonumber \\
 & & - E(\rm {Ti}_{1-\rm {x}}\rm {M}_{\rm {x}}\rm {C} ).
\end{eqnarray}
 The results obtained by evaluating Eq. (\ref{eq:energydiff}) for y=0.25 are given in Fig. \ref{fig:ediff}. Negative values signifies that the carbide may lower its energy by releasing carbon in the form of graphite, i.e. that the right reaction route in Fig. \ref{fig:barrier} is a possible path for decomposition. Generally, graphite formation becomes more favorable by increasing the M concentration. However, in the case of M=V graphite is never formed, which is indicated by a slightly positive $\Delta E$ also for x=1. This is consistent with the V-C phase diagram where a complete miscibility between TiC and VC is observed.\cite{refF} 
It is possible to conclude by looking at Fig. \ref{fig:ediff} that the ability for carbon to be released from the carbide increases with increasing M content. It should also be noted that the concentration for when carbon begins to be released from the carbide depends strongly on the M atom, for Sc and Cr x $\sim$ 0.5  is needed, whereas for Ni, Cu and Zn x $\sim$ 0.3 is needed. In fact, the curves in Fig. \ref{fig:ediff} are directly related to the stability curves in Fig. \ref{fig:formationenthalpy}. The more unfavorable a solid solution is according to Fig. \ref{fig:formationenthalpy}, the greater the driving force to remove C from the structure and reduce the number of M-C bonds. Once more, this is a fact of the decrease in the M-C bond strength with increasing number of valence electrons.
\par
The results of Fig. \ref{fig:ediff} show that it is possible to cause a release of free C from TiC, by dissolving several transition metal species on the metallic sublattice of TiC. This result follows previous theory and experiment, where only the substitution of Al was considered.\cite{wilhelmsson} In the previous work, experimental studies demonstrated an increased rate of the release of free C from Ti$_{1-\rm {x}}$Al$_{\rm {x}}$C$_{\rm {z}}$ when Al was dissolved into the carbide and replacing some of the Ti on the metallic sublattice. This effect was also shown to influence the tribological properties of these carbide films, reducing the friction coefficient by more than 50\% compared to the case of binary TiC.\cite{wilhelmsson} Based on the results in Fig. \ref{fig:ediff} we suggest that a reduced friction coefficient should be expected also for solutions of other metallic species on the Ti sublattice of TiC. It is not unlikely that other transition metal carbides would show a similar pattern.
\par
When regarding the electronic properties of the TMC, it is known that the binary carbides 
 show a more or less rigid band behavior when changing the metal atom.\cite{refEa,refEb} In Fig. \ref{fig:DOStivc} we show calculated density of states (DOS) of Ti$_{1-\rm {x}}$V$_{\rm {x}}$C$_{1-\rm {y}}$ as an illustration of how the electronic structure changes when varying the alloying component for both y=0 and y=0.25.  For TiC  (lowest curve in Fig.~\ref{fig:DOStivc}) the Fermi level is positioned in a region with a low DOS that separates bonding states at lower energies from anti-bonding states at higher energies, which have been formed by hybridization of p and d orbitals originating from the C and Ti atoms respectively.\cite{Grechnev} All the bonding states are therefore occupied while the anti-bonding states at higher energies are empty. By replacing Ti with other metals will effectively move the Fermi level and thereby the occupation of bonding and anti-bonding states depending on the alloying metal. This is the reason for the stability of TiC with respect to the other 3d TMC since all bonding states are occupied while the anti-bonding states are not. In the ternary case, it is clear that the rigid band model can be used to explain the reduced stability of the ternary systems since the DOS is not significantly different when changing the M content from x=0 to x=1, which is illustrated for the case of Ti$_{1-\rm {x}}$V$_{\rm {x}}$C$_{1-\rm {y}}$ in Fig. \ref{fig:DOStivc}. The reduced stability of the ternary phases  are here due to the occupation of anti-bonding pd-hybridized states. Furthermore, the introduction of vacancies on the carbon lattice yields additional features in the DOS which arise from the lowered coordination of the metal atoms in the B1 crystal lattice for this situation.

\section{\label{sec:experimental results}Experimental Results}
The results in the previous section suggest that the addition of a weak carbide-forming metal, substitutionally replacing Ti, should favour the formation of free carbon as graphite. Figure \ref{fig:tinicXPS} shows diffractograms from as-deposited Ti$_{1-\rm {x}}$M$_{\rm {x}}$C$_{1-\rm {y}}$ films with M~=~Ni and Cu. As can be seen, diffraction peaks matching TiC appears in all diffractograms.\cite{Morris} For the Ni-containing series there is a clear shift of the diffraction peaks towards higher diffraction angles, signifying a decrease in lattice parameter from 4.38 {\AA} to 4.32 {\AA} as the Ni-content increases. This is consistent with a substitutional solid solution of Ni in TiC, further details are published elsewhere.\cite{refG1,refG1b} The Cu containing samples do not show the same decrease in lattice parameter. This may be explained by a lesser difference in atomic radii between Ti and Cu.
Furthermore, the theoretical calculations predicts that a solid solution carbide of Ti and Cu 
 should be less stable than the solid solution carbide of Ti and Ni. This is also supported by XRD (see upper right panel of Fig. \ref{fig:tinicXPS}) where the Cu containing samples exhibit a Cu phase, whereas the Ni containing samples synthesized under identical conditions show no Ni phase. 
\par
Experimentally there are two independent steps where the alloying of TiC with a weak carbide forming metal can lead to additional growth of free carbon. Firstly, during the film growth free carbon can be formed together with the carbide. In this case, the as-deposited films will exhibit a carbon phase together with the metastable carbide phase which has a solid solution of the weak carbide forming metal. An increased driving force for graphite formation will then be seen as a larger relative amount of C-C bonds compared to C-M and C-Ti bonds in e.g. an XPS spectrum. Figure \ref{fig:tinicXPS} also shows C1s spectra of the same sample series as was investigated using XRD. As can be seen there are strong peaks at 281.8 eV which can be attributed to C-Ti bonds. This peak has a shoulder which partly can be attributed to C-M bonds and partly to a feature observed in nc-TiC/a-C which is related to interface effects.\cite{Lewin,Lewinb} The peak at 284.2 eV is attributed to C-C bonds, and as can be seen the amount clearly increases as the M content increases, upwards in Fig. \ref{fig:tinicXPS}. Since the total carbon content is approximately constant this means that the carbon content in the carbide phase decreases as a result of the alloying with the weak carbide forming metals Ni and Cu.
\begin{figure}[t]
\begin{center}
\includegraphics[height=6.0cm]{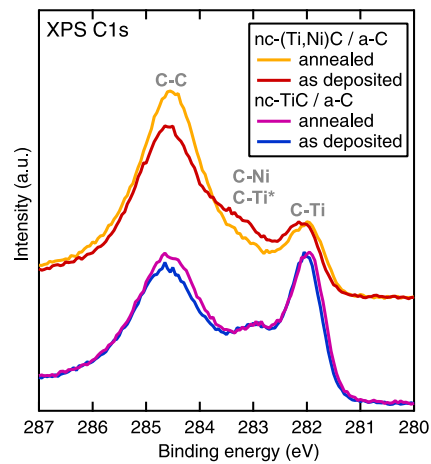}
\caption{(Color online) XPS C1s spectra of annealed and as-deposited thin films. Annealing performed under vacuum at 600$^{\circ}$C for 20 minutes. Ternary nc-(Ti,Ni)C/a-C sample clearly shows an increase in amount of C-phase, in contrast to nc-TiC/a-C sample.}
\label{fig:tinicXPS_annealed}
\end{center}
\end{figure}
Secondly, free carbon can due to the metastable structure of the coating form after film growth during e.g. a high temperature annealing or as a result of external stress in e.g. a tribological contact. Figure \ref{fig:tinicXPS_annealed} shows the result of annealing experiments on a nc-TiC/a-C and a nc-(Ti,Ni)C/a-C thin film. It is clear that the amount of carbon phase is increased more in the case of the annealed ternary sample than in the case of the binary sample. It should be noted that no Ni-phase could be observed by GI-XRD (not shown) after this annealing. Similar results have in previous studies also been attained for Ti-Al-C thin films.\cite{wilhelmsson}
\section{\label{sec:conclusion}Summary and conclusions}
In summary, by using density functional theory calculations, we have identified several transition metal alloyed TiC systems where the introduction of weak carbide forming transition metals create a tendency for the release of carbon from the structure. Energetically, carbon release is seen for all alloying metals that we have studied, except for alloying with~V, but the release sets in for lower concentrations of dissolved metal in the cases of Fe, Co, Ni, Cu, and Zn and for higher concentrations for Sc, Cr, and Mn. The carbon release has been verified experimentally by X-ray photoelectron spectroscopy measurements on thin films for the case of TiC alloyed with Ni and Cu. We note that the out-diffusion of the metal M and a subsequent formation of metal rich phases according to the left reaction route in Fig. \ref{fig:barrier} always will occur if the temperature is high enough to overcome the activation energy barriers for metal diffusion. However, as was pointed out in the introduction, there will be a process window where at lower temperatures the only possible route for decomposition is the release of carbon from the carbide phase.
\par
The electronic structures of the ternary TMC investigated in this study show a rigid band behavior when varying the alloying component from lower to higher values. From the analysis of their electronic structures, we conclude that the reduced stability of the ternary TMC arise from the occupation of anti-bonding states originating from hybridization between C-p and metal-d states. In the case of alloying with Sc, the reduced stability comes from the lesser occupation of the bonding states.
\par
We conclude that for the purpose of creating a material for which the carbon release from the ternary carbide system can be utilized in e.g. a tribological contact, such as in the case of Ti$_{1-\rm {x}}$Al$_{\rm {x}}$C$_{\rm {z}}$ in Ref. \cite{wilhelmsson}, the late transition metals Ni, Cu and Zn would be the best candidates. We also note that due to the similarity of the chemical bonding of 3d TMC and the 4d and 5d TMC, it is very likely that the behavior presented here translates to the 4d and 5d TMC. Furthermore, a number of papers on magnetron sputtering of ternary Ti-M-C films have clearly demonstrated that the addition of weak carbide forming elements such as Fe, Ni, and Pt has a profound effect on the nanocomposite structure of as-deposited films.\cite{refBa,refBb,refBc,refB4,refG1,refG1b, refG2} One of these effects is an increased amount of a matrix phase (free carbon) as demonstrated in Figs.~\ref{fig:tinicXPS} and~\ref{fig:tinicXPS_annealed}. We suggest that this effect during the growth process itself is due to a decreased stability region (with regards to maximum carbon content) of TiC as it is alloyed with a weak carbide forming metal, since this reduces the number of unfavorable M-C bonds.
\section{Acknowledgements}
We acknowledge financial support from the Swedish Research Council, Swedish Foundation for Strategic Research, G{\"o}ran Gustafssons stiftelse and the Carl Trygger Foundation. O. E. is grateful to the European Research Council for support. Valuable discussions with Professor I. A. Abrikosov are also acknowledged. Calculations have been performed at the Swedish national computer centers  SNIC-UPPMAX, SNIC-HPC2N, and SNIC-NSC.

\end{document}